\documentclass{hpl}

\usepackage{graphicx}

\usepackage{amsmath}

\usepackage{xcolor}
\usepackage[normalem]{ulem}

\begin{document}


\shorttitle{Suppression of EMP by Strong Magnetic Fields                                   }
%
\shortauthor{P. V. Heuer et al.}

\title{Suppression of Electromagnetic Pulses from Laser-Target Interactions by Strong Magnetic Fields}
%


\author[1]{P. V. Heuer\corresp{
                       \email{pheu@lle.rochester.edu}}}
\author[1] {J. L. Peebles}
\author[1] {J. R. Davies}
\author[1] {D. H. Barnak}
\author[1] {B. Stanley}
\author[1] {N. Pelepchan}

\author[2] {M. Cufari}
\author[2] {J. A. Frenje}

\author[3] {C. Niemann}
\author[4] {N. A. Rongione} 
\author[3] {C. Constantin}
\author[3] {E. Cisneros} 
\author[3] {P. Pribyl}

\author[5] {H. Sio}
\author[5] {H. Chen}

\address[1]{University of Rochester Laboratory for Laser Energetics, 250 East River Road, Rochester, NY 14623-1299, USA}
\address[2]{Massachusetts Institute of Technology Plasma Science and Fusion Center, Cambridge, Massachusetts 02139, USA }
\address[3]{Department of Physics and Astronomy, University of California, Los Angeles, California, 90024, USA}
\address[4]{Propulsion Science Department, The Aerospace Corporation, Los Angeles, California, 90245, USA}
\address[5]{Lawrence Livermore National Laboratory, Livermore, California, 94550, USA}

\begin{abstract}
Laser-target interactions generate intense electromagnetic pulses (EMP) that can interfere with measurements and damage equipment. In this paper we show that applying a magnetic field to nanosecond pulse laser-target interactions decreases the magnitude of EMP. We demonstrate this effect in two experiments with different geometries (spherical vs. planar), laser intensities (${\sim}10^{13}$ vs. ${\sim} 10^{15}$~W/cm$^2$) and applied field strength (12~T vs. 0.1~T) that both observed suppression of EMP in the ${\sim} 1$~GHz band (by factors of $0.65\times$ and $0.32\times$ respectively). We then observe the opposite effect at high intensities with a picosecond pulse: for planar experiments with laser intensities ${\sim}10^{19}$~W/cm$^2$ and magnetic fields of 6--10~T, the magnitude of EMP is increased by a factor of $1.75\times$. These results provide a benchmark for models of EMP generation, but suggest that magnetic fields are not a viable solution for mitigating EMP on the high intensity laser facilities where it is most damaging. 
\end{abstract}


\keywords{Electromagnetic pulse, EMP, magnetized plasmas}
\maketitle

\section{Introduction}

Laser-target interactions can generate intense electromagnetic pulses (EMPs) in the GHz and THz frequency bands, which can interfere with measurements or damage components of the laser or diagnostics. This risk is partially mitigated with shielding to minimize electromagnetic interference (EMI), but EMP remains a significant concern on many high energy laser facilities and is poised to become an even larger issue on the next generation of ultra-high intensity lasers.

%
%

EMP is generated when electrons are ejected from the target by the laser~\cite{Consoli2020laser, Dubois2014target}. In a simple 1D model, an expanding electron sheath of charge $-2\sigma$ is followed by an ion front of charge $\sigma$, leaving a remaining positive charge $\sigma$ on the target~\cite{Mora2003plasma}. This system can be approximated by an electric dipole with $\sigma$ on the target and $-\sigma$ in the expanding plasma, whose field determines the potential on the target. Above threshold laser intensities (typically $\gtrsim 10^{14}$~W/cm$^2$), ``hot'' electrons are produced near the target surface by laser plasma instabilities (LPI) such as two-plasmon-decay (TPD) and stimulated Raman scattering (SRS). At even higher intensities ($\gtrsim 10^{18}$~W/cm$^2$), additional mechanisms accelerate electrons to even higher energies~\cite{Rusby2024review}. Energetic electrons produced by any of these mechanisms can easily escape the target dipole potential, and in doing so leave an additional positive charge on the target. 

On short (ps) time scales, the accelerating electric dipole moment radiates EMP in the THz band~\citep{Consoli2020laser}. On longer (ns) timescales, the remaining target potential is neutralized by a return current that is typically drawn up the structure supporting the target, typically a thin stalk~\cite{Sinenian2013emperical}. For typical stalk dimensions, this current radiates EMP in the GHz band. The magnitude of the GHz EMP should therefore be proportional to the target potential.

Recent experiments at the Omega Laser Facility~\cite{Cufari2026enhanced} have demonstrated that applying a 10~T magnetic field to a spherical implosion significantly reduces the target potential (measured at the implosion bangtime). As a plasma expands into a background magnetic field, the diamagnetic current $J=\nabla p \times B/B^2$ and the $\vec E \times \vec B$ drift due to the ambipolar field from charge separation expels the magnetic field, forming a diamagnetic cavity and a region of compressed field~\cite{Collette2011structure, Winske2019recalling}. The resulting magnetic field is low at the target surface, then increases at the edge of the cavity, creating a magnetic mirror that reflects hot electrons (generated by LPI) back at the target where they collide and recombine, reducing the target charge. We hypothesized that the magnitude of the GHz EMP should also be reduced, and that the increasing number of electron-target collisions would also lead to an increase in hard x-ray emission on magnetized shots. 

Applying a magnetic field to a target also constrains the expansion of the laser-produced plasma perpendicular to the field~\cite{Harilal2004confinement}. The potential on the target due to the dipole field is proportional to the dipole moment, which in this case is the distance between the target and the expanding plasma. Therefore we expect the target potential due to this dipole moment to be decreased by a magnetic field applied parallel to the target surface. In general, we hypothesize that the target potential is set by this dipole moment in the absence of hot electrons, but is dominated by the charge left by hot electrons in experiments where they are produced. 

In this paper we present results from three sets of experiments. In the first (Sec.~\ref{sec:omega}), on the OMEGA laser, we confirm that GHz EMP generated in spherical implosion experiments in an intensity regime where hot electrons determine the target potential is indeed suppressed by an applied magnetic field. In the second experiment (Sec.~\ref{sec:ucla}, on the Peening laser at the UCLA Phoenix Laser Facility), we show that this effect is also observed at much lower intensities where hot electrons are not generated, with lower field strengths, and in planar geometry. In the third (Sec.~\ref{sec:ep}), on the OMEGA EP laser, we observe that EMP is actually enhanced by an applied magnetic field at high laser intensities (${\sim}10^{19}$~W/cm$^2$) with a correspondingly hot electron population ($\geq 1$~MeV). We conclude in Sec.~\ref{sec:conclusion}.

\section{Spherical implosions\label{sec:omega}}

\begin{figure}[htb]
    \centering
    \includegraphics[width=\columnwidth]{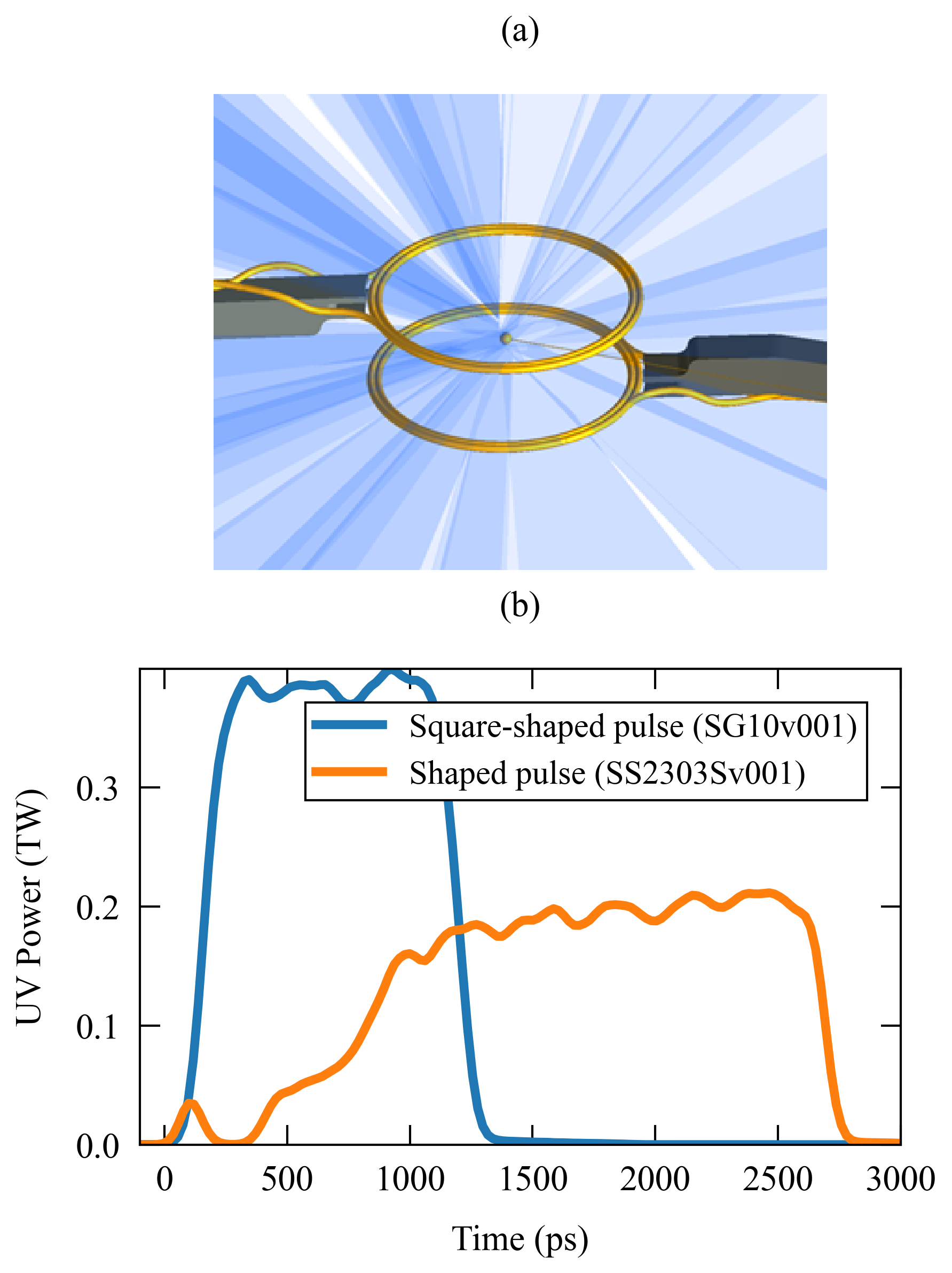}
    \caption{a) 3-D model of the MagSDD platform showing two MIFEDS coils in a Helmholtz configuration around a capsule illuminated by 60 beams. b) Representative examples of the two laser pulse shapes used in the experiments.} 
    \label{fig:omega_setup} 
\end{figure}

\begin{figure}[htb]
    \centering
    \includegraphics[width=0.8\columnwidth]{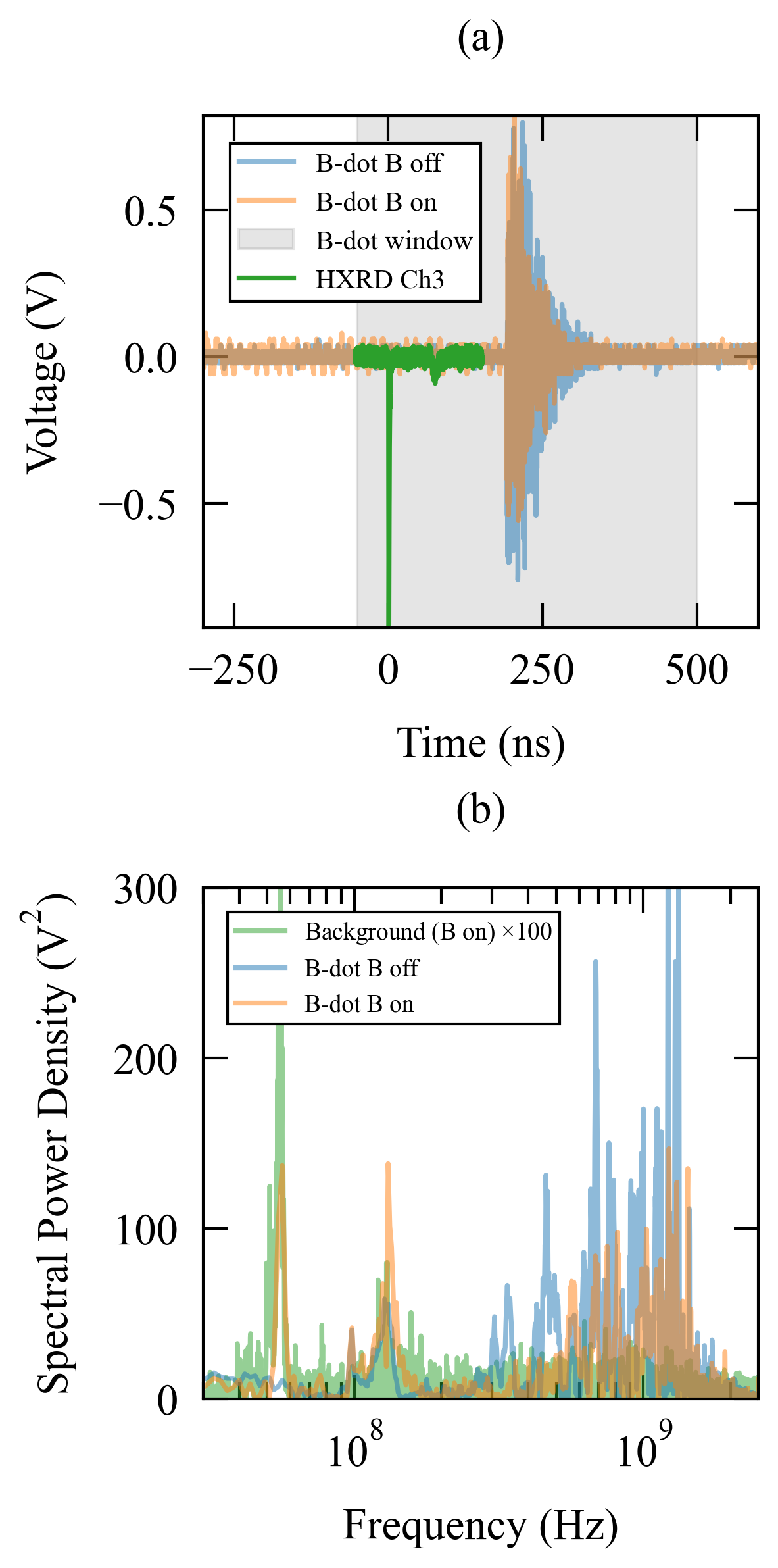}
    \caption{a) Raw B-dot (blue, orange) and hard x-ray diode (green) traces from an unmagnetized shot and a magnetized shot with the MagImp platform. The shaded region shows the time region that was included in the analysis. The x-ray data shown is from the unmagnetized shot: the magnetized signal is indistinguishable on this scale. b) The spectral power density of the B-dot signal within the selected time range from the same two shots. } 
    \label{fig:omega_raw} 
\end{figure}

\begin{figure*}[htb]
    \centering
    \includegraphics[width=1.0\textwidth]{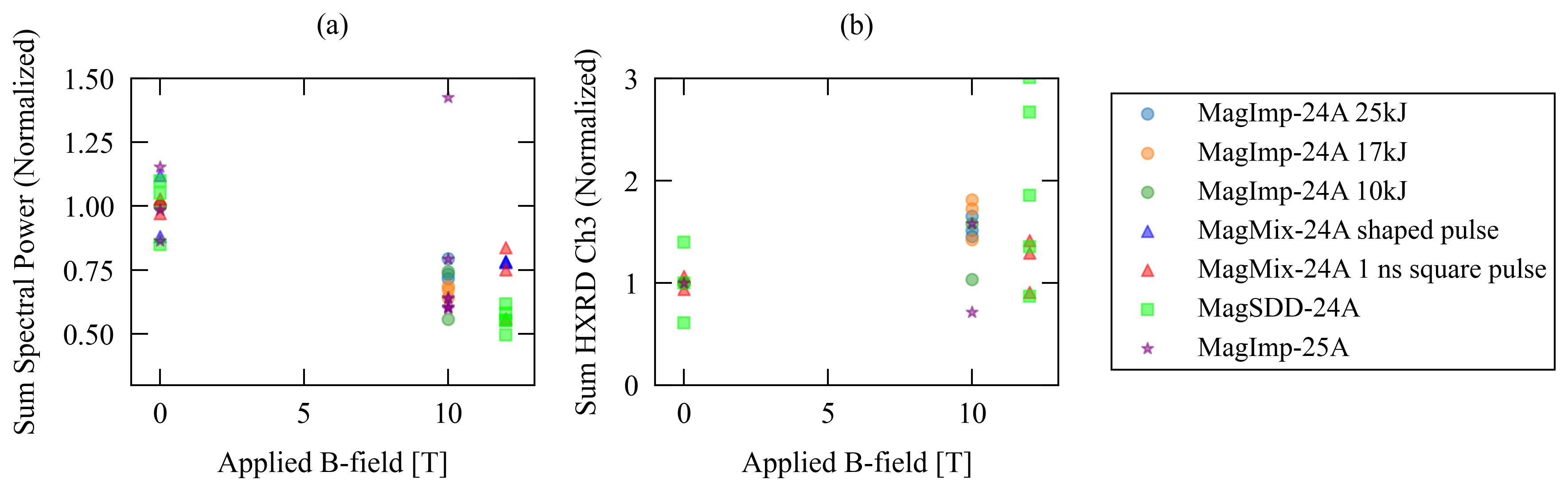}
    \caption{a) Total EMP (summed spectral power of the signal) and b) total summed HXRD channel 3 ($>60$ keV) signal for each shot, each normalized to the mean of the directly comparable unmagnetized shots.} 
    \label{fig:omega_compiled} 
\end{figure*}

EMP measurements were collected by a magnetic flux probe on four campaigns studying magnetized spherical implosions at the Omega Laser Facility (Fig.~\ref{fig:omega_setup}). Two shot days with the MagImp platform used 1--ns square-shaped pulses with total energies ranging from 10 to 25~kJ  (0.5--1 $\times 10^{15}$~W/cm$^2$) to implode ${\sim}850$ $\mu$m outer diameter (OD), 2.5 $\mu$m thick glass shells. Two other shot days with the MagMix and MagSDD platforms used a 20--23~kJ shaped pulse to implode 880 $\mu$m OD, 23.4 $\mu$m thick CH shells. The MagMix shot day tested both 1~ns square pulses and a shaped pulse comprising a picket followed by a flat top drive (Fig.~\ref{fig:omega_setup}b). The MagSDD shot day used the same shaped pulse on all shots. In this regime, hot electrons are generated near the target quarter-critical surface by TPD. In all four experiments, a subset of shots were magnetized using a set of Helmholtz coils pulsed by the magneto-inertial fusion electrical discharge system (MIFEDS)~\citep{Fiksel2015experimental}, providing a peak magnetic field of 10~T for the MagImp platform and 12~T for the MagMix and MagSDD platforms.

Electromagnetic waves generated by experiments on the OMEGA laser were measured by a magnetic flux  or ``B-dot'' probe routinely fielded as part of the EMP monitor (EMPMON) diagnostic. This Prodyn RB-130 B-dot probe is responsive up to 2~GHz and is inserted inside the target chamber 131~cm from the center. Several typical B-dot signals are shown in Fig.~\ref{fig:omega_raw}a. While the laser pulse is only 1~ns long, the EMP signal is longer duration, decaying away exponentially with a time constant of $\tau = 34$~ns, which is a combination of the RC time constant of the stalk discharge current and the resonant modes of the spherical vacuum chamber. 

The power spectrum of the B-dot signal (Fig.~\ref{fig:omega_raw}b) shows peaks near 0.1~GHz ($\lambda {\sim}$ 3 m) and 1~GHz ($\lambda {\sim}$ 30 cm). We take the sum over the power spectrum from 1~MHz to 5~GHz  as a metric for the total EMP energy on a shot. The background spectrum shown is taken from the portion of the magnetized signal prior to the laser shot. Based on the wavelengths, we identify the lower frequency peaks in the EMP spectrum as resonances of the target chamber (3.3~m OD) and the higher frequency peaks as emission from some portion of the target, target stalk, and positioner. In the magnetized shots, the higher frequency emission is suppressed while some lower frequency modes are more strongly excited (Fig.~\ref{fig:omega_raw}b). These peaks are due to the effect of the two MIFEDS units on the resonant modes of the chamber, as the MIFEDS were not inserted on unmagnetized shots. Consequently, these peaks are also observed in the noise spectrum with the MIFEDS inserted (Fig.~\ref{fig:omega_raw}b), while the noise spectrum on the unmagnetized shots without the MIFEDS (not shown) is flat.

To understand the variation in the experimental configurations, we consider the EMP generated on the unmagnetized shots across all experiments. Across all of these shots, the EMP metric varies by a factor of $10\times$. Despite the variation in targets, among the 1--ns square-shaped pulse unmagnetized shots vary by only a factor of $2.5\times$. Comparing shots with identical targets, the shots with the shaped pulse generated an order of magnitude less EMP than those with the 1--ns square-shaped pulse. This is due to the significantly higher peak intensity of the square-shaped pulse compared to the shaped pulse (0.45~TW vs. 0.18~TW), which results in more hot electron production. 

Hard x-rays were diagnosed using the the Omega hard x-ray detector (HXRD)~\citep{Stoeckl2001hard}, which consists of an array of four filtered channels, each with a scintillator coupled to a microchannel plate photomultiplier tube. We focused on the signals from channels~3 and 4, sensitive to $>60$ and $>80$~keV x-rays respectively, which are associated with the stopping of hot electrons in the target. In these implosions x-rays produced by compression are negligible, so we assume that the entire HXRD signal is due to hot electrons and use the integral over the peak in the HXRD signal as a proxy for the number of hot electrons produced. 

In order to compare the effect of the applied magnetic field across campaigns with different targets, laser energies, and pulse shapes, we have divided the data into subsets of directly comparable shots. Within each subset, we normalized the EMP and HXRD metrics to the mean of the unmagnetized shots within that subset. The results are shown in Fig.~\ref{fig:omega_compiled}. By definition, the B=0~T shots cluster around unity. The EMP metric clearly decreases with increasing magnetic field, while the HXRD metric increases. 

The mean and standard deviation of these normalized points for each magnetic field are presented in Table~\ref{table:omega_results}. We note that, within error, the HXRD and EMP metrics are inversely proportional, consistent with the model that electrons colliding with the target produce hard x-rays while simultaneously reducing the target charge and thus the potential energy that generates EMP. 

\begin{table*}[]
\begin{tabular}{|c|c|c|c|c|c|}
\hline
\textbf{B$_0$ (T)} & \textbf{EMP}  & \textbf{\# EMP shots} &  \textbf{HXRD3 ($>60$ keV)}& \textbf{HXRD4 ($>80$ keV)} & \textbf{\# HXRD shots}  \\ \hline
0              & 1.00 ± 0.03    & 13      & 1.00 ± 0.06 & 1.00 ± 0.02  &9       \\ \hline
10             & 0.72 ± 0.05    & 15      & 1.45 ± 0.10 & 1.55 ± 0.10  &10     \\ \hline
12             & 0.66 ± 0.04    & 11      & 1.67 ± 0.26 & 2.28 ± 0.45 &8     \\ \hline
\end{tabular}
\caption{Mean value, standard error, and sample sizes for the EMP measurement (sum spectral power) and  HXRD channel 3 and 4 measurements (summed HXRD signal) normalized to the B=0~T shots within each dataset.\label{table:omega_results}}
\end{table*}

\section{Low intensity planar targets\label{sec:ucla}}

\begin{figure}[htb]
    \centering
    \includegraphics[width=\columnwidth]{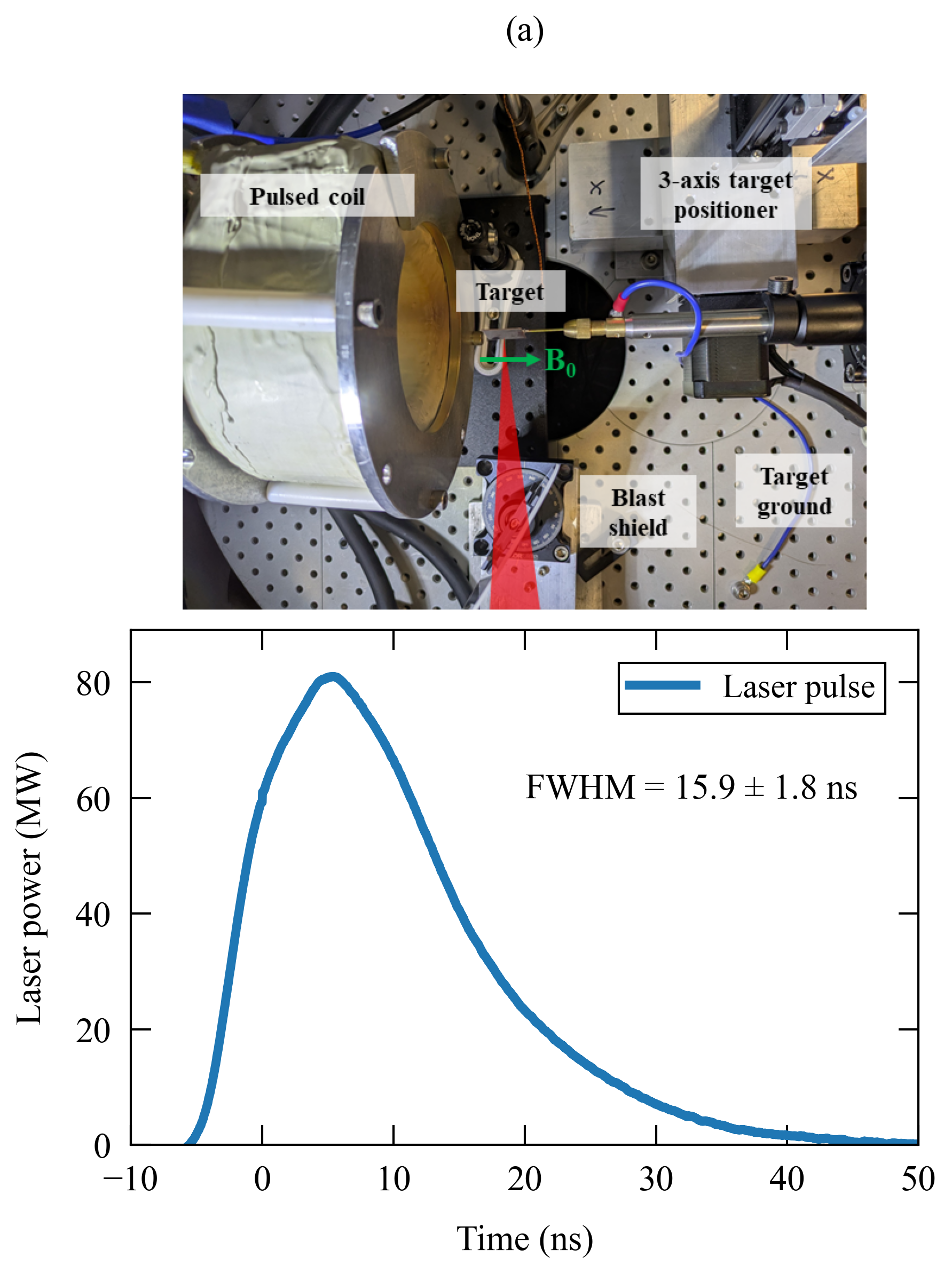}
    \caption{a) A photo of the setup with the magnetic field orientation, target position, and beam path. b) An average of the laser pulse throughout a run, as measured by a pick-off photodiode.} 
    \label{fig:ucla_setup} 
\end{figure}

To investigate the suppression of EMP by magnetic fields at much lower intensities, we conducted a series of shots at the Phoenix Laser Facility at the University of California Los Angeles (UCLA). The experimental setup is shown in Fig.~\ref{fig:ucla_setup}a. The 1053~nm Peening laser~\cite{Hackel1993phase} with a ${\sim} 16$~ns full width half max (FWHM) Gaussian pulse shape (Fig.~\ref{fig:ucla_setup}b) was focused to a 40~$\mu$m FWHM spot on a planar $2 \times 2$~cm 500~$\mu$m thick copper target, tilted by ${\sim} 10^\circ$ to avoid retro-reflection. Shots were taken with 1~J and 4~J of laser energy, $\pm$ 5\%  (5--20 $\times 10^{12}$ W/cm$^2$). This intensity is far below the thresholds for TPD and SRS, and so hot electrons are not generated at the target surface in this experiment.

The target was soldered to a 8~cm long brass stalk, the base of which was connected to chamber ground by a wire. The target was mounted on a 3-axis motion stage and was rastered between shots to present a fresh target surface, keeping the laser focus on the front of the target. A magnetic field of $B_0 \sim 0.1$~T was applied parallel to the target surface by a 40~turn coil pulsed at 2.5~kA and positioned 5~cm from the laser focus on the target. The laser, target drive, and coil pulser were operated at a 0.25 ~Hz repetition rate to collect datasets of several hundred shots. 



\begin{figure}[htb]
    \centering
    \includegraphics[width=0.8\columnwidth]{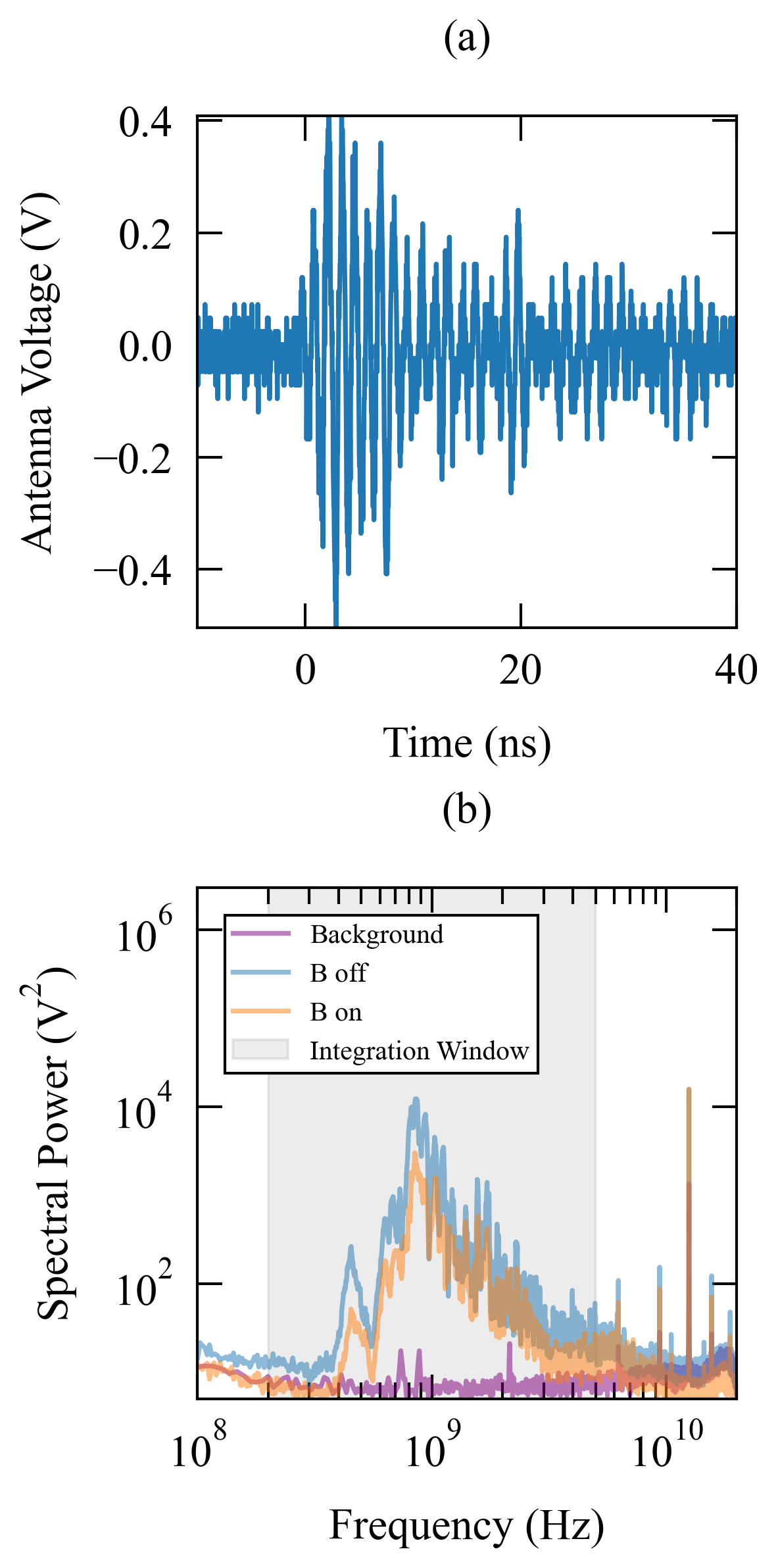}
    \caption{a) A raw single shot antenna trace from the unmagnetized run. b) Median power spectra of the antenna signal across the magnetized and unmagnetized runs.} 
    \label{fig:ucla_raw} 
\end{figure}

\begin{figure}[htb]
    \centering
    \includegraphics[width=0.8\columnwidth]{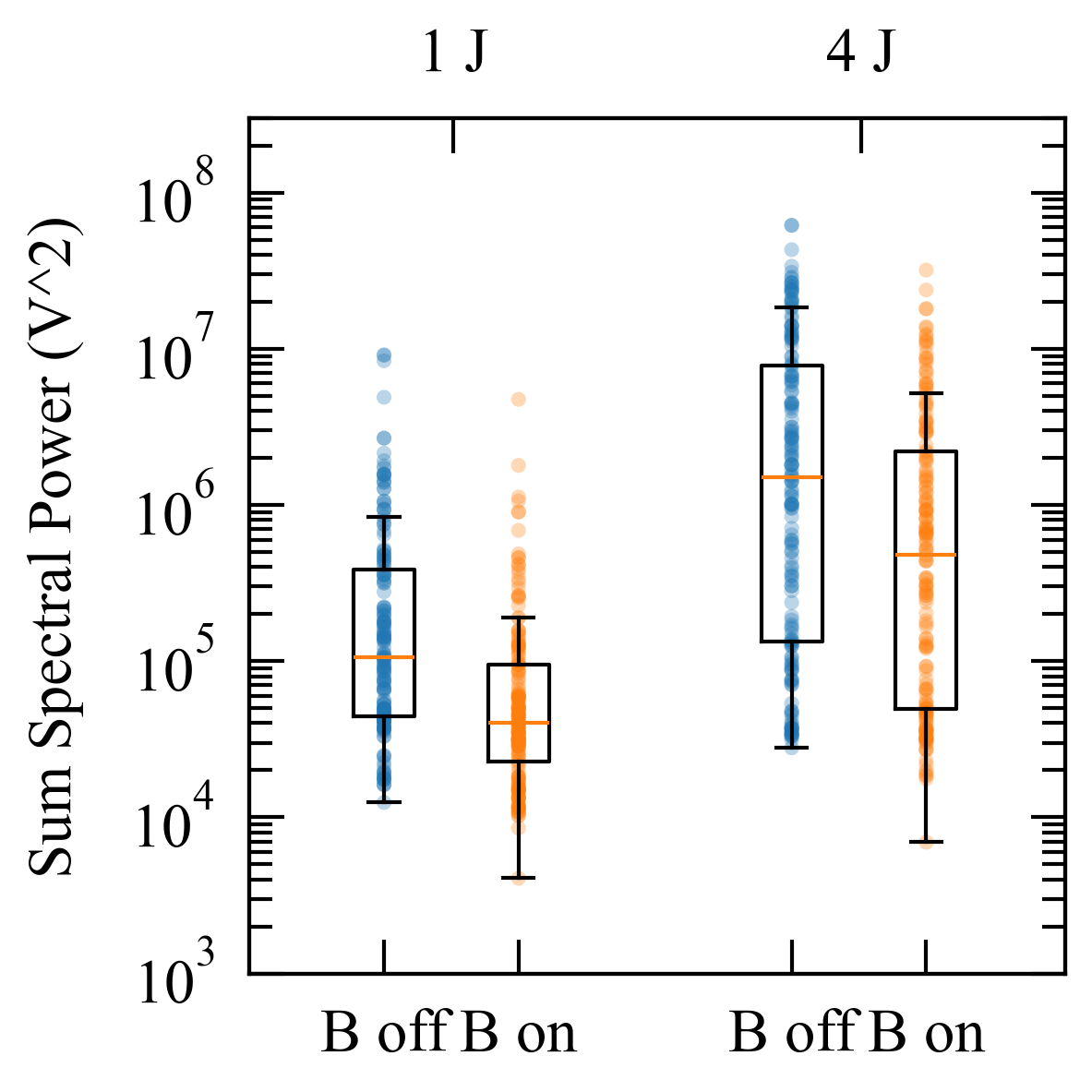}
    \caption{Distribution of summed spectral powers from each shot showing that despite significant variability, there is a clear reduction in EMP power with the magnetic field applied.} 
    \label{fig:ucla_compiled} 
\end{figure}

On each shot, EMP was measured with a Com-Power AH-118 double ridge guide horn antenna which was mounted on the outside of a plastic flange on the target chamber, ${\sim}38$~cm from the center of the chamber, oriented perpendicular to the target stalk. The antenna is nominally sensitive over 0.7-18~GHz, although we measured signal down to $0.4$~GHz. The antenna signal was digitized with a Tektronix DPO72304SX (23~GHz, 50~Gs/s) oscilloscope. Figure~\ref{fig:ucla_raw}a shows a representative antenna signal from a single shot. No attempt is made to correct for the frequency response of the antenna or cables as these are assumed to be identical across all of the shots reported. Unlike in Sec.~\ref{sec:omega}, the EMP signal is observed only while the laser is on.

Figure~\ref{fig:ucla_raw}b shows the median power spectra of the antenna signal during two of the runs. Again two peaks are evident: we identify the peak at ${\sim} 0.5$~GHz with resonances within the 75~cm OD spherical vacuum chamber and the peak at ${\sim} 0.8-2$ GHz as emission from the target and stalk. Again, the background power spectrum is taken from the signal prior to the laser shot. The chamber modes are not evident in the background on this experiment, possibly because the antenna is just outside the chamber. We use the summed spectral power over the gray window marked in Fig.~\ref{fig:ucla_raw}b to represent the total EMP energy as a single scalar. Narrow peaks at $>5$~GHz do not change between shots and are evident in the background spectrum, and so these are attributed to resonances in the antenna circuit. When the magnetic field is applied, there is a clear reduction in magnitude across the spectrum, while its shape is unchanged.

Two sequences of 100 shots each were collected both with and without the applied magnetic field, for a total of 400 shots, at each of two laser energies (1~J and 4~J). A significant amount of shot-to-shot variation was observed on these runs, as evident in the distribution of the summed spectral power for each shot shown in Fig.~\ref{fig:ucla_compiled}. This variation is likely due to fluctuations in the laser intensity on target, which have a strong impact on electron emission and consequently EMP. The Peening laser has no beam smoothing, so hot spots of higher intensity occur randomly on each shot. It is also possible that imperfections in the target surface or the position as the target was rastered modified the intensity. However, both of these effects were stochastic, leading to similar distributions in the magnetized and unmagnetized runs. 

Comparing the distribution of the data in Fig.~\ref{fig:ucla_compiled} shows a clear suppression of the EMP. Using the two-sample Kolmogorov-Smirnov test, the B-on data cannot be consistent with the B-off distribution ($p < 0.1$\%) in either the 1~J or 4~J datasets. The suppression ratio is $0.38\times$ and $0.32\times$ for the 1~J and 4~J datasets respectively.

\section{High intensity planar targets\label{sec:ep}}

\begin{figure}[htb]
    \centering
    \includegraphics[width=0.7\columnwidth]{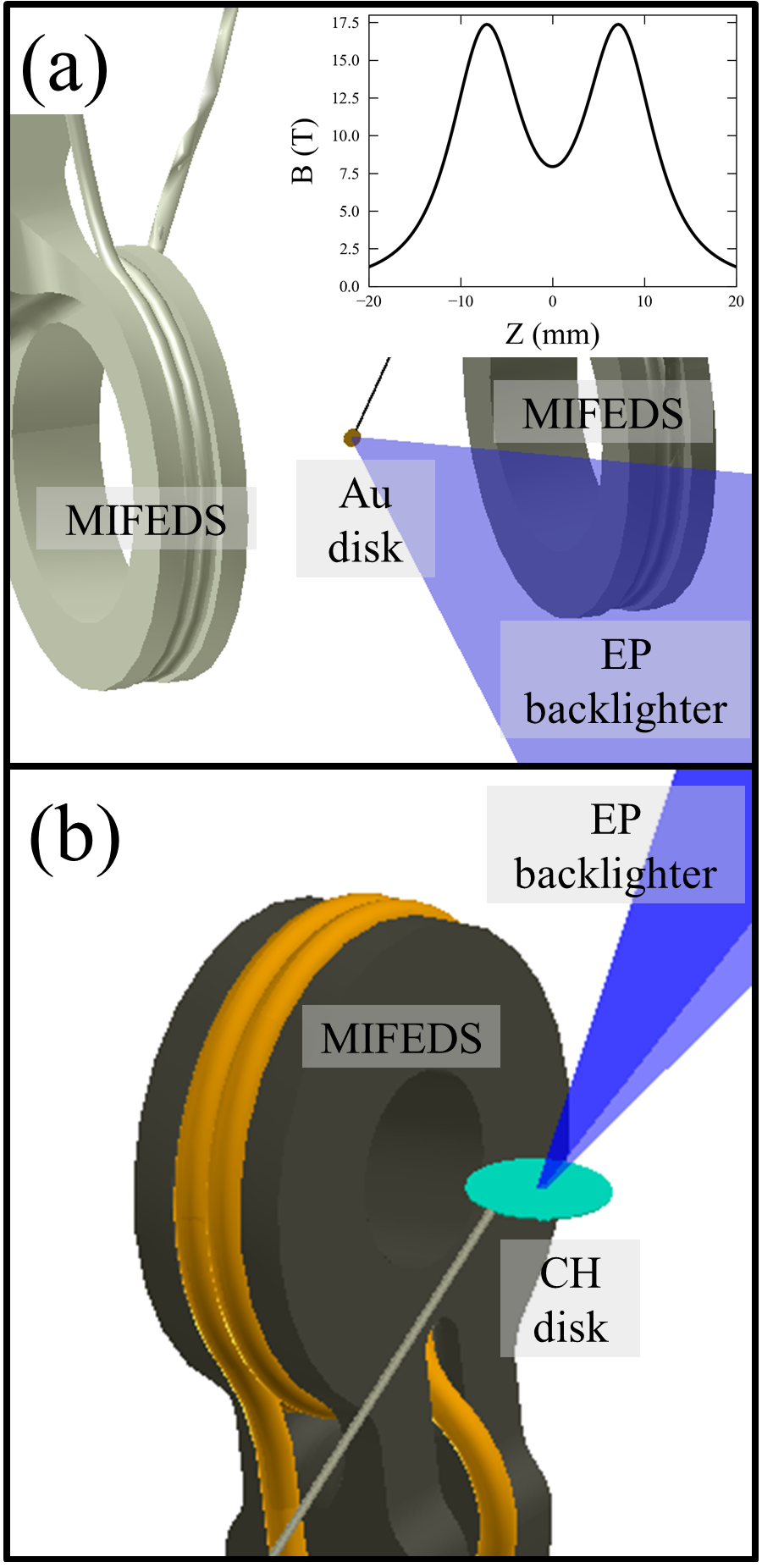}
    \caption{CAD models of the a) PairPlasmaEP and b) BSuppressEMP experiments conducted on OMEGA EP. The inset plot in (a) shows the axial magnetic field profile.} 
    \label{fig:ep_setup} 
\end{figure}

\begin{figure}[htb]
    \centering
    \includegraphics[width=0.8\columnwidth]{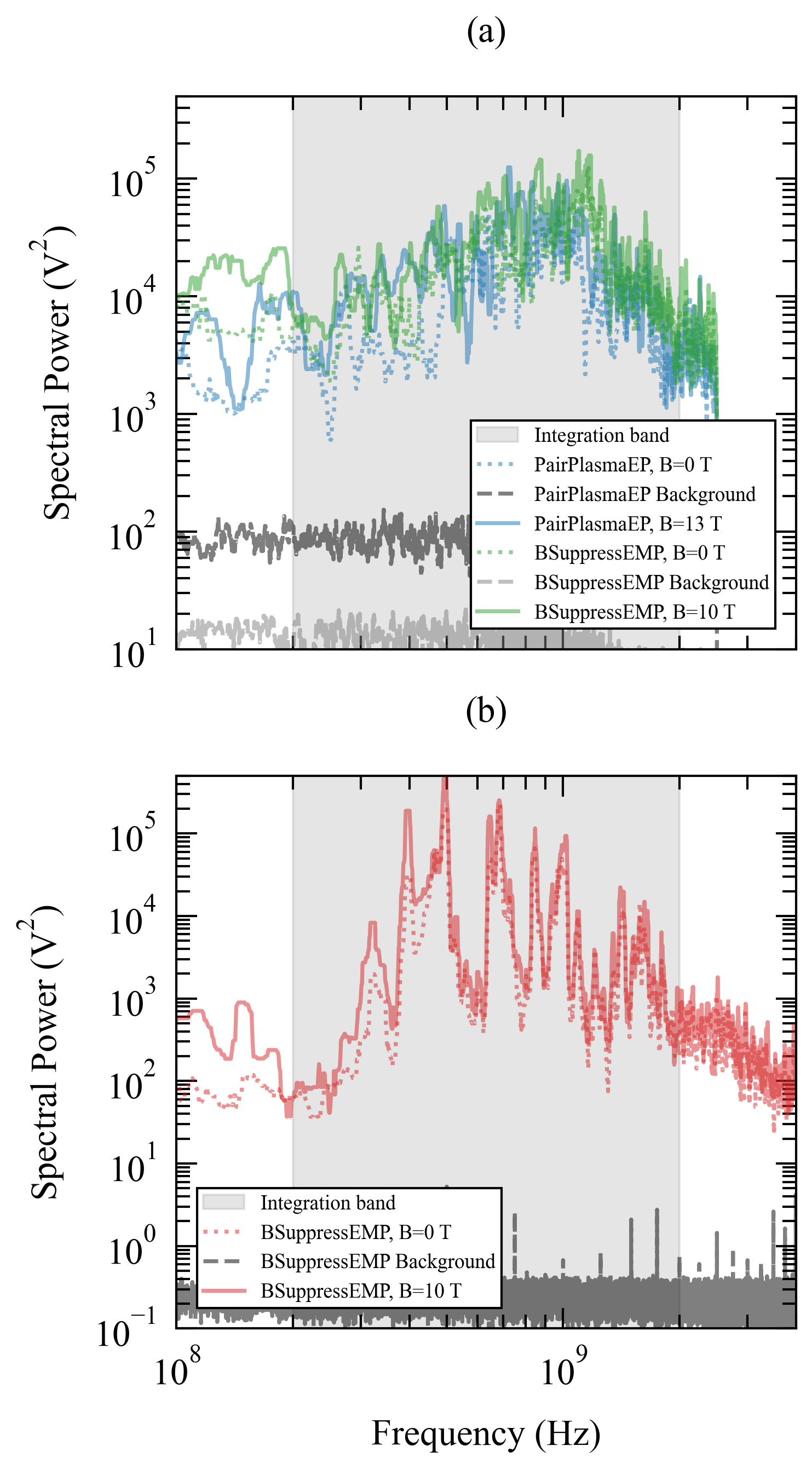}
    \caption{Representative power spectra of the B-dot signals from shots with the OMEGA EP backlighter beam from the a) 131~cm and b) 41~cm B-dot probes. An enhancement in the EMP in the magnetized shots is evident at lower frequencies below 1~GHz.\label{fig:ep_raw}}
\end{figure}

Since the magnitude of EMP is correlated with laser intensity~\cite{Consoli2020laser}, it is a particular threat to high intensity ($>10^{18}$~W/cm$^2$) laser facilities, which can produces hot electrons with temperatures in the range 0.1-10~MeV~\cite{Rusby2024review}. We compared data from two shot days on two distinct platforms on OMEGA EP to asses the impact of magnetic fields on EMP in this regime. 

On the first experiment, named PairPlasmaEP (Fig.~\ref{fig:ep_setup}a), the target was a 500~$\mu$m OD, 20~$\mu$m thick Au disk was shot with the OMEGA EP backlighter beam ($\lambda=1054$~nm) with a 10~ps FWHM pulse and an energy of 880-900~J ($9 \times 10^{18}$~W/cm$^2$). The target was centered between a pair of MIFEDS coils, generating a field of up to 8~T  parallel to the target surface. Two shots each were taken with magnetic field strengths of 5.7~T and 8~T, along with one unmagnetized reference shot. The MIFEDS coils were not inserted on the unmagnetized reference shot. The intended purpose of the experiment was to trap electron positron pairs in this magnetic mirror configuration, and particle tracing simulations and spectrometer measurements suggested that electrons and positrons up to 2.5~MeV are confined for up to 1~ns~\cite{Linden2021confinement}.

On the second experiment, named BSuppressEMP (Fig.~\ref{fig:ep_setup}b), the target was a 2~mm~OD 30~$\mu$m thick CH disk, with a magnetic field of 10~T parallel to the target surface generated by a single MIFEDS coil. The target was again shot with the OMEGA EP backlighter beam with a 10~ps FWHM pulse, but with a lower energy of 200~J ($2 \times 10^{18}$~W/cm$^2$). One magnetized and one unmagnetized shot were collected. The MIFEDS coils were inserted for both shots.

\begin{figure}[htb]
    \centering
    \includegraphics[width=0.8\columnwidth]{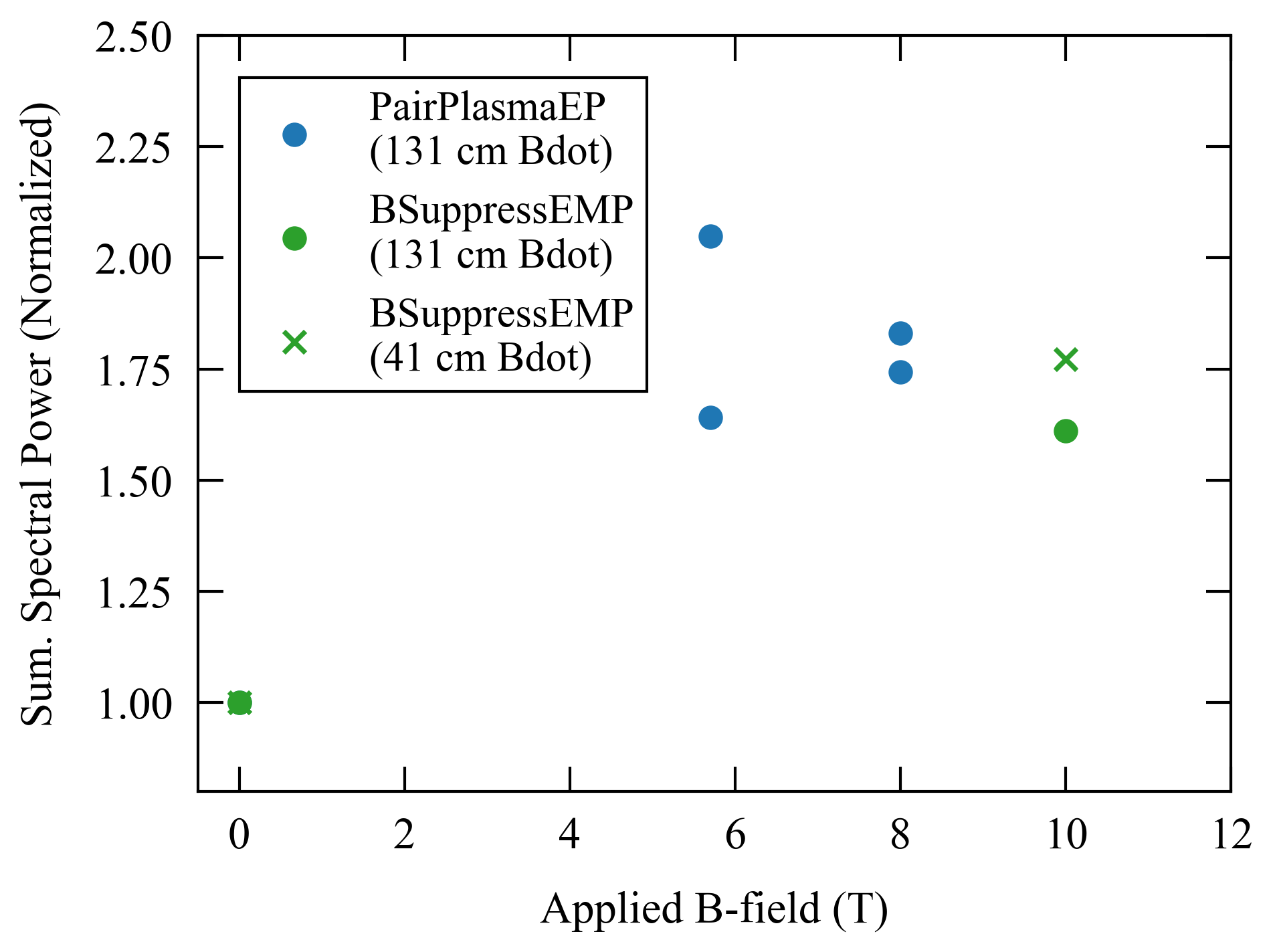}
    \caption{Total EMP (summed spectral power) for each shot, normalized to the mean of the directly comparable unmagnetized shots.} 
    \label{fig:ep_compiled} 
\end{figure}

On both shot days, EMP was measured by a Prodyn RB-130 B-dot probe (responsive to 2~GHz) positioned 131~cm from TCC (identical to the EMP monitor used on OMEGA in Sec.~\ref{sec:omega}). On the BSuppressEMP shot day, a second B-dot probe with a single 1~mm OD loop~\cite{Peebles2022assessment} was inserted 41~cm from chamber center. Example spectra are shown in Fig.~\ref{fig:ep_raw}. As before, to compare data between the two different platforms and probes, the summed spectral power is normalized to that of the corresponding unmagnetized reference shot. Once again the background is extracted from the portion of the signal prior to the shot. In this experiment the EMP is much more intense, requiring additional attenuation on the probes which significantly decreased the background.


The results are summarized in Figure~\ref{fig:ep_compiled}. In stark contrast with the previous sections, we find that applying a magnetic field to these targets increases the magnitude of EMP by ${\sim} 75$\%. We hypothesize that EMP is not mitigated in this regime because electrons that are reflected back to the target now have sufficently high energy that they are unlikely to stop in the target. The range of 1~MeV electrons in gold and polyethylene is 402~$\mu$m and 4.6~$\mu$m respectively~\cite{NIST_PSTAR_ESTAR_ASTAR}, which significantly exceeds the thickness of the target in both OMEGA EP experiments. The mechanism by which the magnetic field actually enhances EMP is less clear. We speculate that the magnetic field may alter the profile of the expanding electrons in a way that increases the dipole moment and/or target potential and therefore the radiated EMP. Additional focused experiments are required to investigate this. However, these preliminary data indicate that the application of magnetic fields may not be an effective strategy for mitigating EMP on intense laser facilities.

\section{Conclusion\label{sec:conclusion}}

Laser-target interactions generate intense electromagnetic pulses (EMP) that can interfere with measurements and damage equipment. In this paper we have presented data from three series of experiments in which we investigate the effect of a magnetic field applied parallel to the target surface on the magnitude of EMP in the ${\sim}1$~GHz band generated by laser-target interactions. In the first experiment, spherical implosions on the OMEGA laser at ${\sim}10^{15}$~W/cm$^2$, EMP was suppressed by a factor of $0.65$--$0.72\times$ by a 10--12~T applied field. On this experiment, an increase in hard x-rays was also observed with the applied field, consistent with more electrons returning to the target and neutralizing the target potential. In the second experiment in planar geometry using the Peening laser at the UCLA Phoenix Laser Laboratory with an intensity of $5\times10^{13}$~W/cm$^2$, a suppression of $0.32\times$ was observed with only a 0.1~T applied field. 

In a third series of experiments, gold and plastic targets were magnetized to 6--10~T and shot on OMEGA EP at ${\sim}10^{19}$~W/cm$^2$. In this case, the applied magnetic field enhanced the EMP emitted by a factor of 1.75$\times$. We hypothesize that the magnetic field does not suppress EMP in this regime because hot electrons with energies above a few 100~keV are unlikely to stop and deposit their charge in thin targets. Future experiments are necessary to test this hypothesis, and to understand the enhancement of EMP in this regime. 

It is possible that the applied magnetic field changes the directionality of the EMP emission rather than its magnitude, but that the magnitude appears to be reduced at the limited number of detectors available. However, our data suggests that this is not the case. If expanding electrons were confined along the magnetic field we would expect the dipole moment to increase parallel to the field, increasing emission perpendicular to the field. In each of our experiments, the probes or antenna are positioned approximately perpendicular to the field, and so should see an increase in magnitude rather than the decrease observed. On OMEGA, the correlation of the decreased EMP with the increase in hard x-rays suggests that the EMP magnitude is being suppressed. In the UCLA experiment the spherical chamber modes, which should effectively average over the EMP emission in all directions, are suppressed by the same amount as the higher frequency EMP. On OMEGA EP, the 131~cm and 41~cm B-dot probes were fielded at different angles to the field, but observed similar suppression of EMP. We conclude that only suppression of the magnitude of the EMP by the applied field is consistent with these results.

\medskip
\noindent\textbf{Acknowledgements}

This material is based upon work supported by the Department of Energy [National Nuclear Security Administration] University of Rochester “National Inertial Confinement Fusion Program” under Award Number(s) DE-NA0004144. The work in section~\ref{sec:ucla} was supported by the U.S. Department of Energy’s (DOE) Office of Science (SC) Fusion Energy Sciences (FES) program under DE-SC0024549: the LaserNetUS initiative at the Phoenix Laser Laboratory.

This report was prepared as an account of work sponsored by an agency of the U.S. Government. Neither the U.S. Government nor any agency thereof, nor any of their employees, makes any warranty, express or implied, or assumes any legal liability or responsibility for the accuracy, completeness, or usefulness of any information, apparatus, product, or process disclosed, or represents that its use would not infringe privately owned rights. Reference herein to any specific commercial product, process, or service by trade name, trademark, manufacturer, or otherwise does not necessarily constitute or imply its endorsement, recommendation, or favoring by the U.S. Government or any agency thereof. The views and opinions of authors expressed herein do not necessarily state or reflect those of the U.S. Government or any agency thereof.

\medskip

\end{document}